\documentclass[twocolumn]{aastex62}

\usepackage[utf8]{inputenc}
\DeclareUnicodeCharacter{2212}{-}
\usepackage{graphicx}
\graphicspath{{./}{Figures}}
\usepackage{amsmath}
\usepackage{ulem}
\usepackage{breqn}
\usepackage{afterpage}
\usepackage{enumitem}
\usepackage{xcolor}
\usepackage{float}
\usepackage{xspace} 
\usepackage{url}
\usepackage{fontawesome}
\setenumerate{itemsep=0mm}
\usepackage{lineno}
\usepackage{wrapfig}
\usepackage{placeins}

\newcommand{\ocen}{\ensuremath{\omega {\rm Cen}}\xspace}

\newcommand{\mbh}{\ensuremath{{\rm M_{\rm BH}}}\xspace}
\newcommand{\mstar}{\ensuremath{{\rm M_\star}}\xspace}
\newcommand{\msun}{\ensuremath{{\rm M_\odot}}\xspace}
\newcommand{\mnsc}{\ensuremath{{\rm M_{\rm NSC}}}\xspace}
\newcommand{\z}{\ensuremath{z}\xspace}
\newcommand{\edd}{Eddington\xspace}

\usepackage{amsmath}
\usepackage{amssymb}
\usepackage{xspace}
\usepackage{xifthen}
\usepackage{eso-pic}

\definecolor{forestgreen}{HTML}{228B22}
\definecolor{urlblue}{HTML}{000000}

\mathchardef\mhyphen="2D

\newlength{\dhatheight}

\newcommand{\unit}[1]{\ensuremath{\mathrm{\,#1}}\xspace}

\newcommand{\km}{\unit{km}}
\newcommand{\kms}{\km \second^{-1}}

\newcommand{\second}{\unit{s}}

\newcommand{\bandvar}[2][]{%
  \ifthenelse{\isempty{#1}}{\var{#2}}{\var{#2\_#1}}%
}

\newcommand{\var}[1]{\ensuremath{\texttt{\MakeUppercase{#1}}}\xspace}

\providecommand\physrep{\ref@jnl{Phys.~Rep.}}%
\providecommand\apjs{\ref@jnl{ApJS}}%
\providecommand{\jcap}{\ref@jnl{JCAP}}%

\begin{document}

\title{\textbf{
Black Hole Scaling Relations in the Dwarf-galaxy Regime \\ with \textit{Gaia}-Sausage/Enceladus and $\omega$Centauri
}}
%


\correspondingauthor{Guilherme Limberg}
\email{limberg@uchicago.edu}
\author[0000-0002-9269-8287]{Guilherme~Limberg}
\affiliation{Kavli Institute for Cosmological Physics, University of Chicago, 5640 S Ellis Avenue, Chicago, IL 60637, USA}
\affiliation{Universidade de S\~ao Paulo, IAG, Departamento de Astronomia, SP 05508-090, S\~ao Paulo, Brasil}

 \vspace{-6mm}

\begin{abstract}

The discovery of fast moving stars in Milky Way's most massive globular cluster, $\omega$Centauri (\ocen), has provided strong evidence for an intermediate-mass black hole (IMBH) inside of it. However, \ocen is known to be the stripped nuclear star cluster (NSC) of an ancient, now-destroyed, dwarf galaxy. The best candidate to be the original host progenitor of \ocen is the tidally disrupted dwarf \textit{Gaia}-Sausage/Enceladus (GSE), a former Milky Way satellite as massive as the Large Magellanic Cloud. I compare \ocen/GSE with other central BH hosts and place it within the broader context of BH-galaxy (co)evolution. The IMBH of \ocen/GSE follows the scaling relation between central BH mass and host stellar mass (\mbh--\mstar) extrapolated from local massive galaxies ($\mstar \gtrsim 10^{10}\,\msun$). Therefore, the IMBH of \ocen/GSE suggests that this relation extends to the dwarf-galaxy regime. I verify that \ocen (GSE), as well as other NSCs with candidate IMBHs and ultracompact dwarf galaxies, also follow the \mbh--$\sigma_\star$ relation with stellar velocity dispersion. Under the assumption of a direct collapse BH, \ocen/GSE's IMBH would require a low initial mass ($\lesssim$10,000\,\msun) and almost no accretion over $\sim$3\,Gyr, which could be the extreme opposite of high-$z$ galaxies with overmassive BHs such as GN-z11. If \ocen/GSE's IMBH formed from a Population III supernova remnant, then it could indicate that both light and heavy seeding mechanisms of central BH formation are at play. Other stripped NSCs and dwarf galaxies could help further populate the \mbh--\mstar and \mbh--$\sigma_\star$ relations in the low-mass regime and constraint IMBH demographics and their formation channels.
\end{abstract}



\section{Introduction} \label{sec:intro}

Understanding the early assembly of supermassive black holes (BHs) and their coevolution with their host galaxies is a major goal in Astrophysics \citep{Kormendy2013blackHoles, Inayoshi2020smbh, Fan2023quasars}. Given a certain seeding mechanism, either Population III supernovae explosions or direct collapse \citep[][]{Volonteri2008seeds, Volonteri2010bh_formation}, BHs should populate the entire range from stellar mass BHs ($\mbh \sim10\,\msun$; e.g., \citealt{GaiaBH3}) to supermassive ones (${>}10^6\,\msun$) inhabiting in the centers of massive galaxies, including the Milky Way \citep[MW;][]{EckartGenzel1997sgrA*, Ghez1998sgrA*}. However, the population of intermediate-mass BHs (IMBHs; $100\lesssim \mbh/\msun < 10^5$) expected to reside in dwarf galaxies (stellar mass $\mstar \lesssim 10^9\,\msun$) has remained elusive \citep[][]{Reines2022dwarfBHs} with few detections \citep{nguyen2019ngc205, woo2019ngc4395}.

Just recently, \citet{Haberle2024imbhOmegaCen} has provided strong evidence for the existence of an IMBH ($\mbh \gtrsim 8200\,\msun$) in $\omega$Centauri (\ocen), the most massive MW globular cluster \citep[e.g.,][]{Baumgardt2018massGCs}; see \citet[][]{noyola2008ocen} for an earlier proposition. These authors identified a collection of fast moving stars 
with tangential velocities well above the cluster escape velocity, which require the presence of a central IMBH to remain bound, but see \citet{Banares-Hernandez2024_oCen_PulsarTiming} for other possibilities. 
However, it has been known for more than 20\,years that \ocen is 
rather the stripped nuclear star cluster (NSC) of a former dwarf galaxy that has been fully tidally disrupted by the MW \citep[][]{Lee1999omeCen, BekkiFreeman2003omeCen}. 

With the advent of \textit{Gaia} \citep[][]{GaiaMission}, multiple disrupted dwarfs have been identified in the MW's halo \citep[e.g.,][]{naidu2020}, including \textit{Gaia}-Sausage/Enceladus \citep[GSE;][]{belokurov2018, Haywood2018, helmi2018}, with an \mstar in between the Small and Large Magellanic Clouds (SMC and LMC). This merger event represents the last major merger experienced by the MW $\sim$10\,Gyr ago (redshift $z \sim 2$) with a mass ratio in between 1:3 and 1:5 \citep[][]{Naidu2021simulations, Amarante2022gsehalos}. 
\ocen has similar kinematics to GSE's stellar population \citep[][]{massari2019, Callingham2022gcs} and follows the relation between NSC mass (\mnsc) and host \mstar with GSE \citep[][]{Limberg2022gse}. Hence, 
GSE is the best candidate, out of known accreted dwarfs, to be the original host galaxy of \ocen. Therefore, the discovery of a central BH in \ocen provides the unique opportunity to extend well-known scaling relations, such as between \mbh and host \mstar \citep[][]{ReinesVolonteri2015mbh}, into the regime of IMBHs and dwarf galaxies.

In this \textit{Letter}, I contextualize the IMBH inside \ocen, hence GSE, with the \mbh--\mstar relation and data at different redshifts. The most important result is that the \ocen/GSE system follows the local \mbh--\mstar relation extrapolated from massive galaxies ($\mstar \gtrsim 10\,\msun$), extending it into the IMBH/dwarf-galaxy regime. I also verify that \ocen, as well as other NSCs with candidate IMBHs and compact galaxies, follows the relation between \mbh and stellar velocity dispersion $\sigma_\star$. I calculate possible BH growth trajectories to show that, if \ocen was formed via heavy seeding from direct collapse, it must have had a very low initial mass and experienced almost no accretion during its lifetime until the GSE merger with the MW. Lastly, I compare other candidate MW NSCs with the \mnsc--\mstar relation and discuss its implications. \citet[][]{PlanckCollab2020} cosmology is adopted throughout.

\begin{figure*}[pt!]
\centering
\includegraphics[width=2.1\columnwidth]{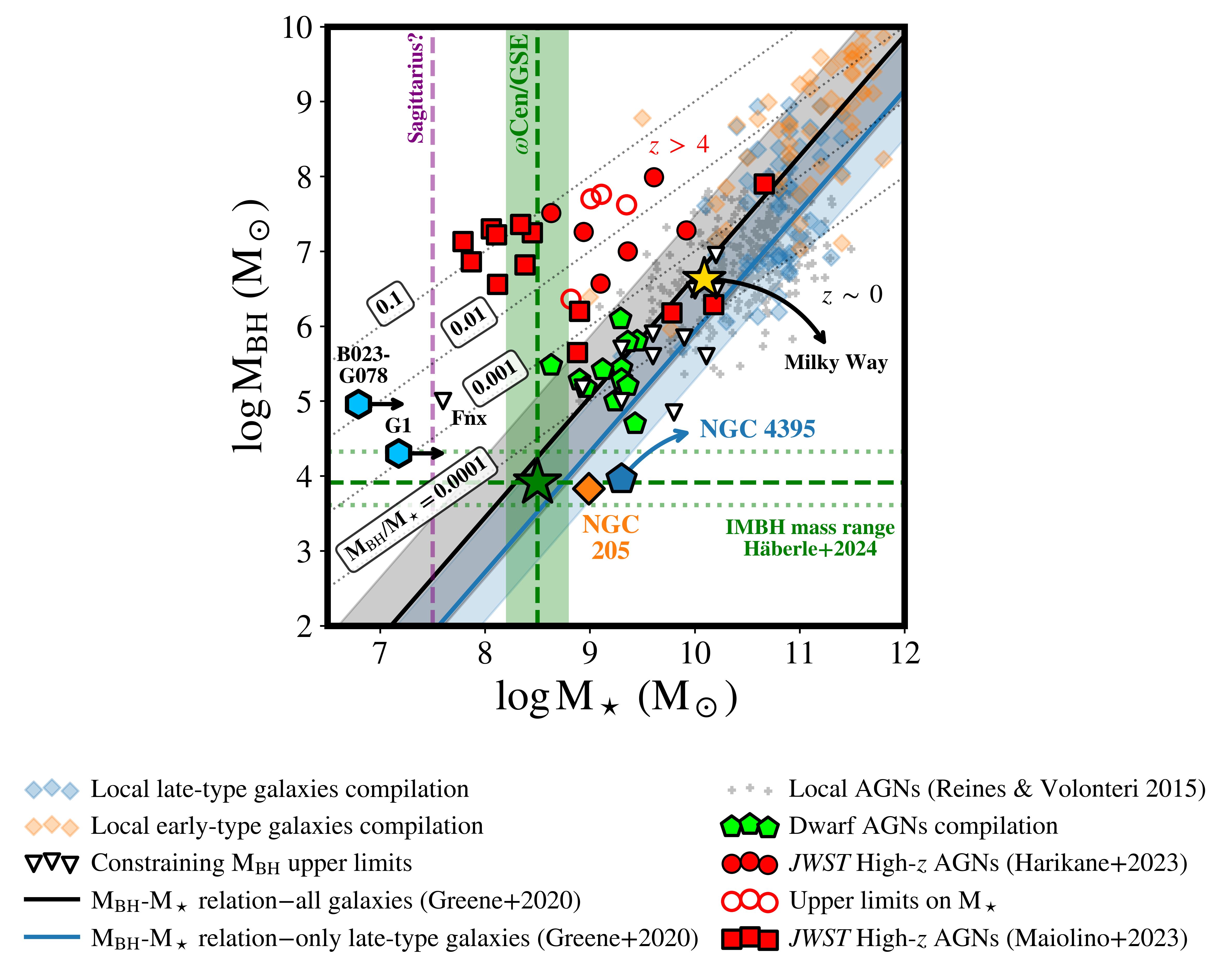}
\caption{\mbh--\mstar. Local ($z \sim 0$) regular (non-AGN) galaxies are shown as blue and orange diamonds for
late- and early-type galaxies, respectively \citep[][]{Greene2020imbh}. White triangles with black edges are constraining \mbh upper limits derived from stellar dynamics \citep[also compiled by][]{Greene2020imbh}, including Fornax dSph \citep[Fnx;][]{Jardel2012fornaxBH}. MW is in yellow \citep[][]{Genzel2010bhs}. Local AGNs are shown as the gray `$+$' signs \citep[][]{ReinesVolonteri2015mbh}. Green pentagons show a compilation of dwarf AGNs \citep[also by][]{ReinesVolonteri2015mbh}. Red squares and circles are high-$z$ AGNs \citep[$z > 4$;][respectively]{Maiolino2023jwst, harikane2023jwst}. The \mbh--\mstar relations from \citet[][]{Greene2020imbh} for all galaxies and late-type only are shown in black ($\pm$0.81\,dex scatter around the mean) and blue ($\pm$0.65\,dex), respectively. The \ocen/GSE, where $\log{(\mstar/\msun)} = 8.5$, is plotted as the green star symbol \citep[][]{Callingham2022gcs}. The green $\pm$0.3\,dex stripe covers the typical \mstar reported in literature for GSE (see text). The IMBH mass range allowed by the \citet[][]{Haberle2024imbhOmegaCen} analysis is displayed as the green dotted lines. Other IMBH detections, namely NGC 205 \citep[orange diamond with black edge;][]{nguyen2019ngc205} and NGC 4395 \citep[blue pentagon;][]{woo2019ngc4395}, are also plotted. The $\log{(\mstar/\msun)} = 7.5$ of Sagittarius dSph is represented by the purple dashed line \citep[][]{Callingham2022gcs}. M31 stellar clusters with candidate IMBHs are the blue hexagons with black edges \citep{Gebhardt2002g1, Pechetti2022imbh}. For these M31 candidates, I adopt cluster masses as lower limits to their host \mstar. Black dotted lines represent different values for the \mbh/\mstar ratio. 
\label{fig:relat}}
\end{figure*}

\section{Data} \label{sec:data}

At redshift $z \sim 0$, \mbh and \mstar values for regular galaxies (non-active galactic nuclei, AGNs) are from \citet[][]{Greene2020imbh}, which is augmented with respect to the compilation by \citet[][]{Kormendy2013blackHoles}. These galaxies have dynamical \mbh measurements and self-consistent \mstar estimates \citep[as in][]{Bell2003_Mstar}. I also adopt scaling relations from \citet[][]{Greene2020imbh}, which accounts for low-mass galaxies and (IM)BHs by including \mbh upper limits in their fits \citep[e.g.,][]{NeumayerWalcher2012} as well as few detections in this regime \citep{nguyen2019ngc205, woo2019ngc4395}. Although not used for the scaling relations, Figure \ref{fig:relat} also shows the local sample of broad-line AGNs from \citet[][
]{ReinesVolonteri2015mbh}. For reference, I also plot these authors' compilation of dwarf AGNs (their table 3). 
The high-\z spectroscopic sample consists of \textit{JWST} data for AGNs at $z > 4$ \citep[][
]{Maiolino2023jwst, harikane2023jwst}. For the exploration of the \mbh--$\sigma_\star$ relation (Figure \ref{fig:m-sigma}), I additionally consider dwarf AGNs with $\sigma_\star$ measurements from \citet[][]{Baldassare2020dwarfAGNs} and compact galaxies with detected central massive BHs \citep[][]{Seth2014ucdBH, Afanasiev2018ucdBH, Ahn2018ucdBH}. 

The mass range for \ocen's IMBH is from \citet[][]{Haberle2024imbhOmegaCen}; $4100 \leq \mbh/\msun \leq 21{,}100$. Note that these are possible lower limits on the IMBH mass. Nevertheless, these authors rule out $\mbh$ values above $50{,}000\,\msun$. The qualitative insights in this \textit{Letter} do not depend on the exact \mbh value adopted within this range. The \mnsc and host \mstar are compiled by \citet[][]{Neumayer2020reviewNSC}. I also utilize these authors' \mnsc--\mstar relation, close to $\mnsc \propto \mstar^{1/2}$. All MW globular cluster masses for the candidate NSCs are from \citet[][2021 revision]{Baumgardt2018massGCs}; the \ocen mass is $\mnsc \equiv 4 \times 10^6\,\msun$. The $\sigma_\star = 22.6\,\kms$ value for \ocen is taken from the recent analysis by \citet[][]{Pechetti2024oCen}, which is consistent with previous work \citep[$\sim$23\,$\kms$,][]{noyola2008ocen}.

All \mstar values for disrupted dwarf galaxies are from \citet[][]{Callingham2022gcs}. These authors use globular cluster counts to estimate total mass \citep[][]{BurkertForbes2020}. Then, \mstar is found through the relation with halo mass \citep[][]{Behroozi2019_UNIVERSEmachine}. As a sanity check, I calculate \mstar for Sagittarius dwarf spheroidal (dSph) using this galaxy's and the Sun's absolute $V$-band magnitude \citep[$M_V = -13.27$ and ${+4.81}$, respectively;][]{Majewski2003, willmer2018solar_absM}. I find $7.4 \leq \log{(\mstar/\msun)} < 7.6$ with mass-to-light ratios between 1.5 and 2.2 \citep[e.g.,][]{Kirby2013} whereas \citet[][]{Callingham2022gcs} estimate $\log{(\mstar/\msun)} = 7.5\pm0.4$. For GSE, \citet[][]{Callingham2022gcs} finds $\log{(\mstar/\msun)} = 8.5\pm0.3$, which is well within the realm of literature values \citep[][]{Lane2023_GSEmass}. The green stripe in Figure \ref{fig:relat} covers this $\pm$0.3\,dex interval. For GSE's accretion redshift, I adopt $z = 2$ (look-back time of $\sim$10.5\,Gyr) as inferred from stellar age distributions \citep[][]{gallart2019, Bonaca2020}. Again, this value is well accepted to be the timing of the merger \citep{Naidu2021simulations, Amarante2022gsehalos}.

\section{Discussion} \label{sec:discussion}

\subsection{The IMBH of GSE inside \texorpdfstring{\ocen}x extends the local \texorpdfstring{\mbh}x--\texorpdfstring{\mstar}x relation} \label{subsec:relation}

The main takeaway from Figure \ref{fig:relat} is that the IMBH of \ocen/GSE resides on top of the \mbh--\mstar scaling relations from \citet[][]{Greene2020imbh}; see text below for possible variations. Since GSE analogs in cosmological hydrodynamical simulations are found to be ubiquitously star-forming systems leading to gas-rich mergers with their MW-like hosts \citep[e.g.,][]{Bignone2019, Grand2020}, we consider \mbh--\mstar relations for both all galaxies and late-type only samples. Different possible \mstar and/or \mbh values for \ocen/GSE would not change this qualitative conclusion given the scatters (shaded areas around the black and blue relation lines). The consequence is that S/LMC-mass galaxies ($10^8 < \mstar/\msun \lesssim 10^9$, similar to GSE) might follow a natural extension of high-mass systems ($\mstar \gtrsim 10^{10}\,\msun$) in the \mbh--\mstar relation; note the claimed IMBHs in NGC 205 \citep[][]{nguyen2019ngc205} and NGC 4395 \citep[][]{woo2019ngc4395} also plotted in Figure \ref{fig:relat}. Therefore, other MW and/or M31 satellites and stripped NSCs could be promising targets to further expand IMBH demographics in the dwarf-galaxy regime. Observational challenges are many \citep[see][]{Reines2022dwarfBHs}, but this approach would allow us to further contextualize IMBHs within the broader picture of galaxy-BH coevolution
.

\textit{JWST} has revealed a surprising excess of overmassive BHs in faint AGNs at $z \gtrsim 4$ (red symbols in Figures \ref{fig:relat} and \ref{fig:zspec}), far exceeding the number density expected from canonical quasar luminosity functions \citep[e.g.,][]{Kokorev2024lrd}. These AGNs host central BHs where $\mbh/\mstar \gtrsim 0.01$, well above the expected from the local \mbh--\mstar relation \citep[see][]{Pacucci2023high-z}. Many of these high-$z$ AGNs have similar \mstar to GSE, but central BHs of order 100--1000$\times$ more massive than \ocen's IMBH. To reconcile the fast growth of these overmassive BHs in the early universe with the IMBH of \ocen/GSE, similar host \mstar at different $z$, might be challenging for BH--galaxy coevolution theory, but useful for supermassive BH growth and seeding models (Section \ref{subsec:seeding}). 


Many previous works, including some aforementioned ones, have derived the \mbh--\mstar scaling relation from various samples and found discrepant shapes for it \citep[][and \citealt{Pacucci2023high-z} for very high $z$ only]{Kormendy2013blackHoles, ReinesVolonteri2015mbh, Greene2020imbh}
. 
These empirical relations basically all converge in the high-mass regime ($10^{10} < \mstar/\msun \lesssim 10^{12}$), which is expected given that all of them are derived from data for massive galaxies. However, their slopes can be dramatically different, ranging from ${\sim}\mbh \propto \mstar$ to $\mbh \propto \mstar^3$ \citep[see][]{shankar2016_MbhMstar}. 
Hence, when approaching the IMBH/dwarf-galaxy regime, some of these fits severally under predict, by factors of $\sim$1000$\times$, the \mbh of \ocen/GSE. These might simply mean that these relations should not be extrapolated below $\mstar \approx 10^{10}\,\msun$. Having said that, we basically have constraints neither on the scatter around \mbh nor on the BH occupation fraction for GSE-mass (S/LMC-mass) galaxies with so few IMBH detections: NGC 205 \citep[][]{nguyen2019ngc205}, NGC 4395 \citep[][]{woo2019ngc4395}, and now \ocen/GSE \citep[][]{Haberle2024imbhOmegaCen}. Therefore, although the exact choice of \mbh--\mstar relation might change the exact statement that \ocen/GSE follows an extrapolation from massive galaxies, we can \textit{now} extend it to the IMBH/dwarf-galaxy regime. 


Apart from the broader implications to the \mbh--\mstar relation for dwarf galaxies, the formation/evolution of a GSE-mass system with its central IMBH is also interesting to explore. It would be important to understand how the central dynamics of dwarf galaxies and star formation in them respond to the presence of an IMBH within the specific $\mbh/\mstar$ ratios allowed for \ocen/GSE \citep[e.g.,][]{Koudmani2021bhFeedback}. Likewise, the evolutionary pathway of GSE-mass galaxies with central (IM)BHs could be useful for refining feedback models since AGN activity is expected to regulate the \mbh--\mstar relation itself \citep[][]{DiMatteo2005AGNfeedback}, although that might not be the case for dwarf galaxies \citep[][]{Sharma2020bhFeedback}. The accretion history of the IMBH in \ocen/GSE could even tell us about how stellar feedback regulates the early growth of central BHs in dwarf galaxies \citep[][]{Angles-Alcazar2017bhSNfeedback, Habouzit2017bhSNfeedback}. Moreover, \ocen/GSE's IMBH might hold clues to the role of environment (isolated versus satellite dwarf) since merger activity could play a role in providing the necessary conditions for BH growth \citep[][]{Volonteri2008seeds}. \ocen/GSE could even be explored as a laboratory for IMBH ``incubation'' in NSCs, where the host would provide the necessary gas supply for a stellar mass BHs to grow to the intermediate-mass status \citep{Natarajan2021_BHincubation}.

\subsection{\texorpdfstring{\ocen}x (GSE) in the \texorpdfstring{\mbh}x--\texorpdfstring{$\sigma_\star$}x relation} \label{subsec:m-sigma}

\begin{figure}[pt!]
\centering
\includegraphics[width=1.0\columnwidth]{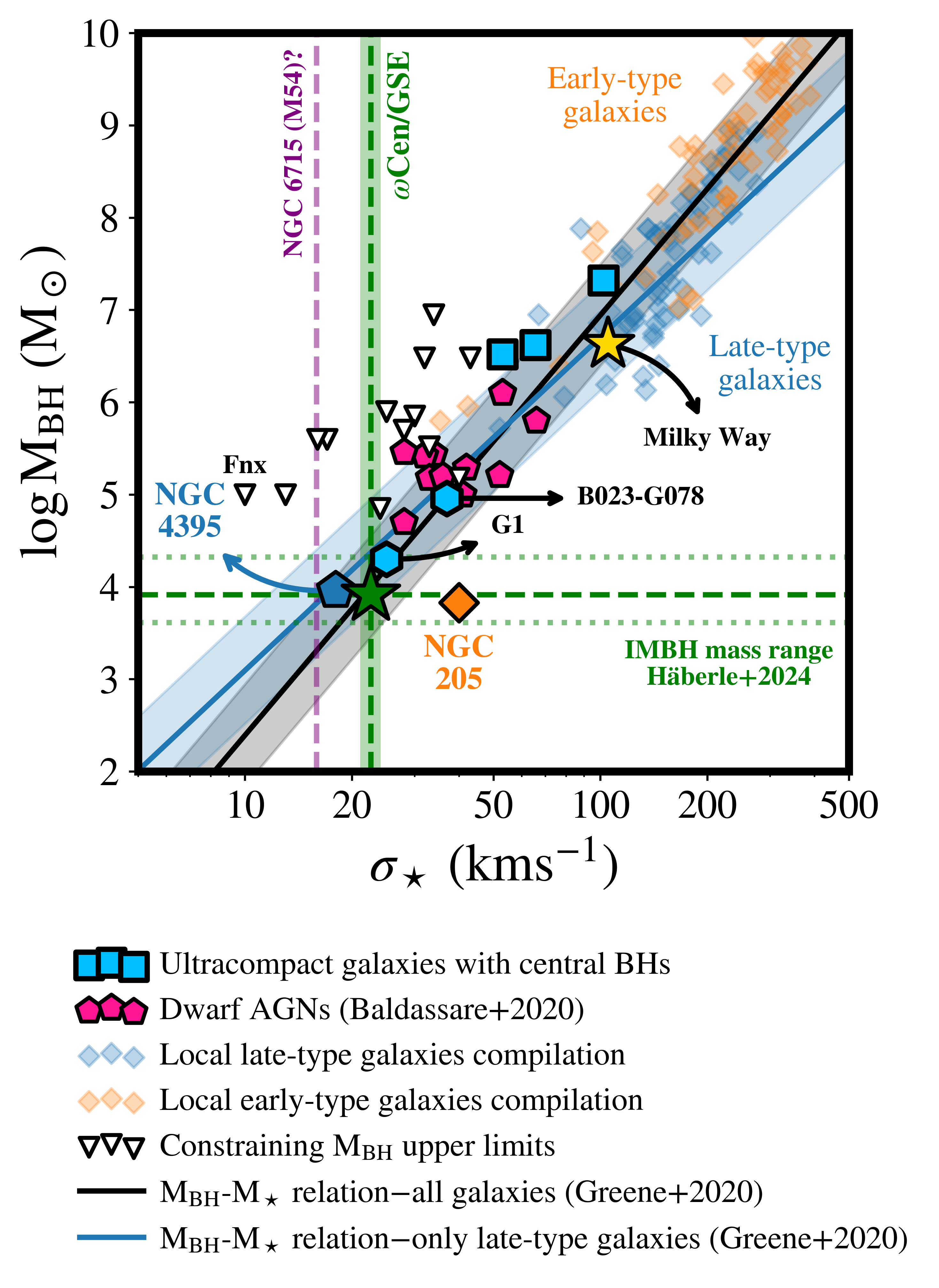}
\caption{\mbh--$\sigma_\star$. Symbols and colors follow the scheme of Figure \ref{fig:relat}; blue and orange diamonds are late- and early-type galaxies, respectively \citep[][]{Greene2020imbh}, triangles are constraining \mbh upper limits \citep[also][]{Greene2020imbh}, MW in yellow \citep[][]{Genzel2010bhs}, M31 stellar clusters with candidate IMBHs as the blue hexagons \citep[][]{Gebhardt2002g1, Pechetti2022imbh}, and \ocen (GSE) in dark green ($\sigma_\star$ from \citealt{Pechetti2024oCen}). Blue squares are UCDs with massive BHs detected (see text). Pink pentagons are dwarf AGNs with $\sigma_\star$ measurements from \citet[][]{Baldassare2020dwarfAGNs}. Other IMBH detections, NGC 205 \citep[orange diamond with black edge,][]{nguyen2019ngc205} and NGC 4395 \citep[blue pentagon,][]{woo2019ngc4395}, are also shown. The $\sigma_\star$ value of NGC 6715/M54 is the purple dashed line \citep[][]{Alfaro-Cuello2020m54sigma}. The \mbh--$\sigma_\star$ relations from \citet[][]{Greene2020imbh} are plotted in black ($\pm$0.55\,dex scatter) and blue ($\pm$0.58\,dex) for all galaxies and late-type only, respectively.
\label{fig:m-sigma}}
\end{figure}

I also explore the location of \ocen/GSE's IMBH in the \mbh--$\sigma_\star$ relation (Figure \ref{fig:m-sigma}). Previous works have attempted to constraint the slope of this relation toward the low-mass regime, in particular leveraging dwarf AGNs \citep[][]{Xiao2011M-sigma, Schutte2019dwarfAGNs, Baldassare2020dwarfAGNs}. The \mbh--$\sigma_\star$ relations derived by \citet[][]{Greene2020imbh}, and adopted here, also extend to dwarf galaxies/IMBHs by incorporating constraining \mbh upper limits in their fits, mostly from \citet{NeumayerWalcher2012} and including Fornax dSph \citep[][]{Jardel2012fornaxBH}. For \ocen, we take the recent $\sigma_\star$ value from \citet[][]{Pechetti2024oCen}. We refer the reader to the \mbh--$\sigma_\star$ relation plot by \citet[][their figure 10]{Barth2009ngc3621} for an earlier version of this exercise, including both \ocen \citep[data from][]{noyola2008ocen} and G1 \citep[][]{Gebhardt2002g1}, another stellar cluster with an IMBH candidate known at the time (see Section \ref{subsec:nsc}).

\ocen and its IMBH reside on top of the \citeauthor{Greene2020imbh}'s (\citeyear{Greene2020imbh}) \mbh--$\sigma_\star$ relation for all galaxies (within scatter for late-type only; Figure \ref{fig:m-sigma}). I verify that M31 stellar clusters with candidate IMBHs, the aforementioned G1 \citep[][]{Gebhardt2002g1} and B023-G078 \citep[][]{Pechetti2022imbh}, also closely follow the relation. Another class of stellar systems related to NSCs with detected central massive BHs \citep[][]{Seth2014ucdBH, Afanasiev2018ucdBH, Ahn2018ucdBH} are the so-called ``ultracompact'' dwarf galaxies \citep[UCDs, $10^6 \lesssim \mstar/\msun \leq 10^8$,][]{Drinkwater2003ucd}. Akin to \ocen, these UCDs are thought to be remnants of stripped NSCs \citep[][]{wang2023ucd}, and are usually found in dense environments such as galaxy clusters \citep[e.g.,][and references therein]{Liu2020ucd}. Figure \ref{fig:m-sigma} shows that UCDs also appear to follow the \mbh--$\sigma_\star$ relation, perhaps reinforcing their evolutionary connection to NSCs. Although samples are still quite small, the emerging broad picture takeaway might be that, since both NSCs and UCDs alike follow the \mbh--$\sigma_\star$ relation, the $\sigma_\star$ in these low-mass systems can be compared to those in the bulges of local massive galaxies.

\subsection{Implications for BH seeding} \label{subsec:seeding}

The main classes of BH seeding prescriptions are ``light seeds'' (Population III supernovae remnants) and ``heavy seeds'' (direct collapse); see \citet[][]{Volonteri2010bh_formation}. The latter might be favored by the high-$z$ population of overmassive BHs discovered by \textit{JWST}. 
For example, galaxy GN-z11 \citep[][]{oesch2016gn-z11}, similar $\log{(\mstar/\msun)} = 8.9$ to GSE, has a central BH of $\mbh = 1.5 \times 10^6\,\msun$ already at $z=10.6$ \citep[][]{Maiolino2024gn-z11}. On the other hand, the IMBH of \ocen/GSE has an estimated \mbh quite close to the minimum value expected by heavy seeding models \citep[e.g.,][${\approx}10^4\,\msun$]{Volonteri2008seeds}. Hence, the IMBH in \ocen/GSE could, in principle, establish a lower limit to the mass distribution of heavy seeds \citep[][]{Reines2022dwarfBHs}.

\begin{figure}[pt!]
\centering
\includegraphics[width=1.0\columnwidth]{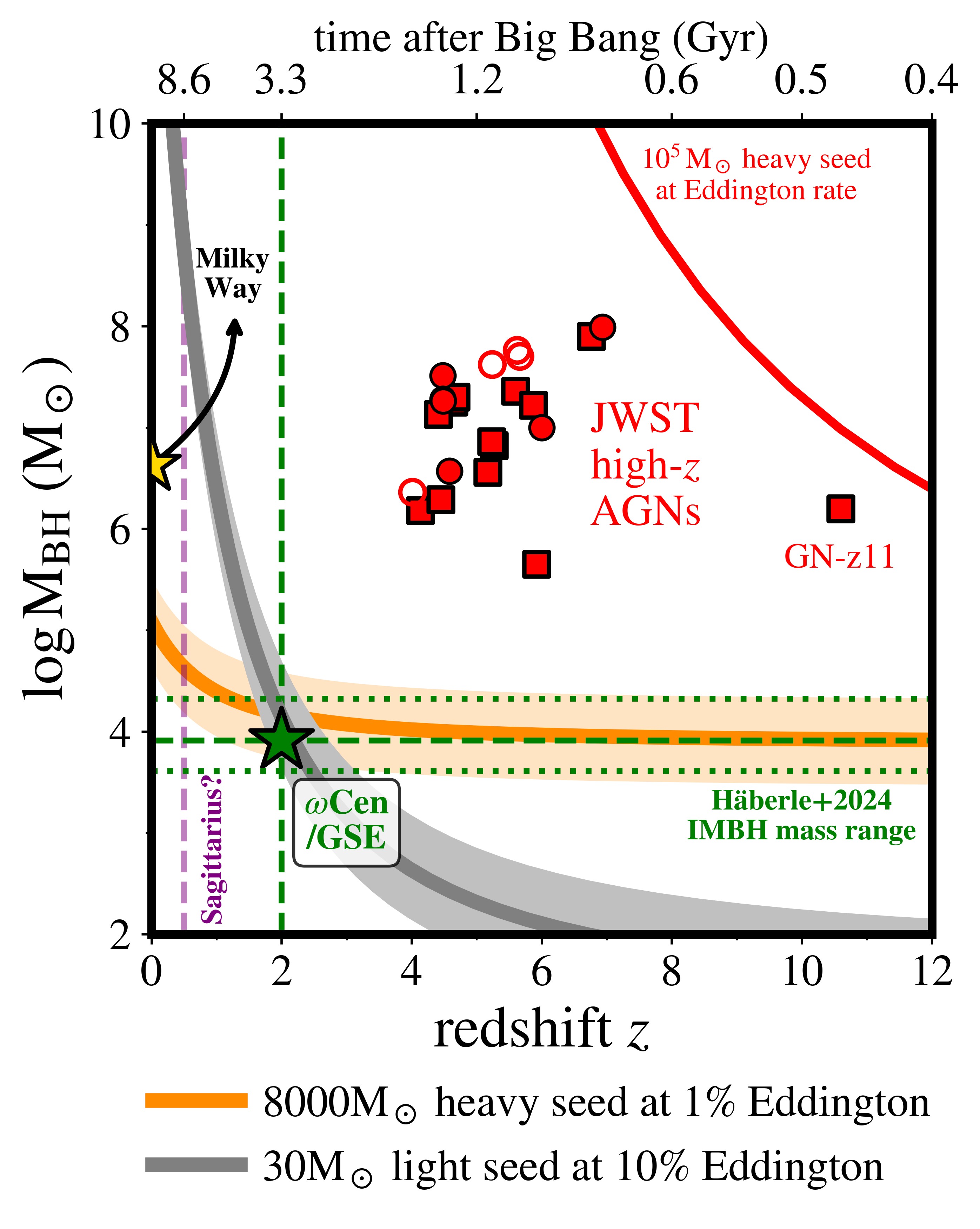}
\caption{\mbh--redshift. The corresponding time since Big Bang is calculated with \citet[][]{PlanckCollab2020} cosmology (upper axis). Symbols and colors follow the same scheme of Figure \ref{fig:relat}; red squares and circles for high-$z$ AGNs \citep[][respectively]{Maiolino2023jwst, harikane2023jwst}, Miky Way in yellow \citep[][]{Genzel2010bhs}, and \ocen/GSE in dark green. Purple line represents a previous, perhaps first, pericenter of Sagittarius dSph \citep[][]{Ruiz-Lara2020sag}. BH growth tracks calculated as in \citet[][see text]{Fan2023quasars} are shown as the orange (heavy seed) and gray (light seed) lines. The range of initial masses are 10--100\,\msun for light seeds and 5000--20{,}000\,\msun for heavy ones. 
\label{fig:zspec}}
\end{figure}

The IMBH in \ocen must have stopped accretion at the time of the merger due to gas removal by the MW via ram-pressure and/or tidal stripping, as evidenced by the fast quenching of GSE analogs in simulations \citep[][]{Bignone2019, Grand2020}, hence being ``frozen'' at $z \sim 2$. Indeed, this scenario could be consistent with \ocen's IMBH exquisitely weak accretion rate implied by upper limits in X-ray luminosity \citep[][]{Haggard2013oCenXray, Tremou2018gcsRadio}. I consider this fact to estimate possible BH growth trajectories for \ocen/GSE (similar to \citealt{Pacucci2023high-z} for high-$z$ galaxies). For this exercise, I use the same formalism as \citet[][their equations 7 through 9]{Fan2023quasars}. Under the hypothesis of heavy seeding, one potentially important implication is that the IMBH of \ocen/GSE must have experienced extremely low accretion rate throughout its \textit{entire} $\sim$3\,Gyr lifetime. Assuming an 8000\,\msun initial mass at $z=18$ and a constant accretion rate, the IMBH of \ocen/GSE should have grown at only $\sim$1\% of the Eddington luminosity. If I adopt a light seed of 30\,\msun (similar to the \mbh of the recently discovered Gaia BH3; \citealt{GaiaBH3}), the \mbh of \ocen/GSE's IMBH can be achieved at a fixed 10\% Eddington rate (Figure \ref{fig:zspec}).

This analysis of BH growth tracks 
should be regarded as initial exploration, but it points to a scenario where the IMBH of \ocen/GSE could represent the complete opposite extreme from GN-z11 and its high-$z$ counterparts \citep[][]{harikane2023jwst, Maiolino2023jwst} in terms of possible BH growth histories with heavy seeding. These very high-$z$ AGNs with overmassive central BHs likely require a combination of both high initial \mbh and accretion rate (sometimes $\sim$50,000\,\msun and $\gtrsim$80\% \edd, \citealt{Pacucci2023high-z}), perhaps including super-\edd episodes \citep[][]{Maiolino2024gn-z11}; see the recent detection by \citealt{Suh2024superEdd}
. On the other hand, \ocen/GSE's IMBH would have a low initial mass and experience almost no accretion. 

If Population III supernovae seeding is allowed, one interpretation could be that both mechanisms are at play, light and heavy seeding simultaneously, hence creating a true physical distinction between supermassive and intermediate-mass central BHs. 
This reinforces the need to hunt for additional IMBHs to constrain both their properties and their host dwarf galaxies'. Sagittarius dSph might be interesting for testing that hypothesis since the local \mbh--\mstar relation predicts that the mass of a central BH in that dwarf galaxy would be of $\lesssim$1000\,\msun \citep[][Figure \ref{fig:relat}]{Greene2020imbh}. 

\subsection{Other accreted NSCs in the MW} \label{subsec:nsc}

If the discovery of an IMBH in \ocen is confirmed, the next step would be to extend the search to other NSCs with the obvious candidate being NGC 6751 (M54), the NSC of Sagittarius dSph galaxy \citep[][]{Bellazzini2008m54}. However, M54 is ${\sim}5\times$ farther from the Sun than \ocen 
and is located behind the Galactic bulge, so stars are fainter, more difficult to resolve, and the field is plagued with MW foreground. Moreover, M54 still resides within the dark-matter halo of Sagittarius and this galaxy is undergoing severe tidal stripping by the MW \citep[][]{Majewski2003}, which might affect its central dynamics.

\begin{figure}[pt!]
\centering
\includegraphics[width=1.0\columnwidth]{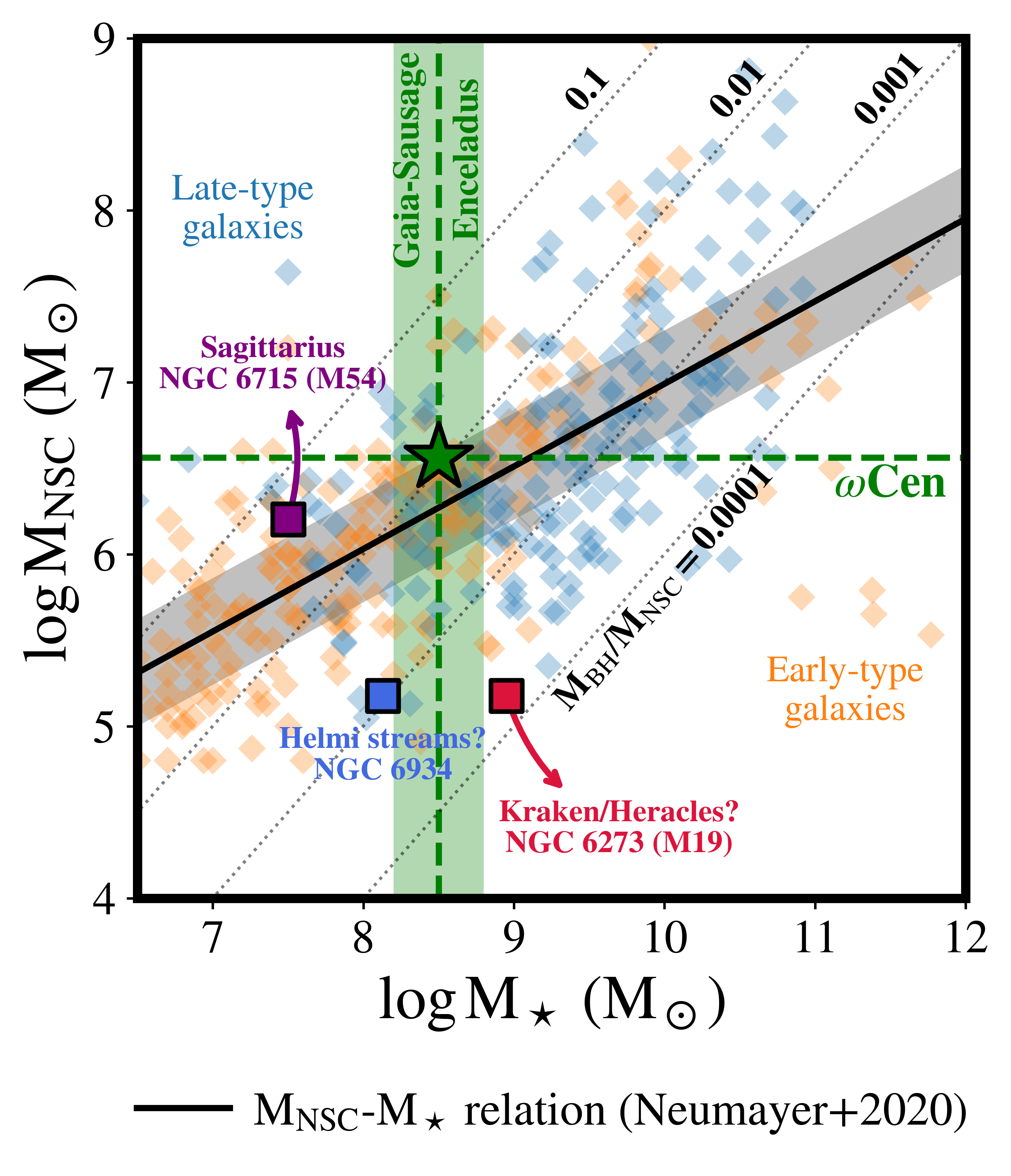}
\caption{\mnsc--\mstar. Blue and orange diamonds represent late- and early-type galaxies, respectively. The \mnsc--\mstar relation is the black line ($\pm$0.6\,dex scatter as the shaded region; \citealt{Neumayer2020reviewNSC}). Squares are different MW candidate NSCs and host galaxies (see text); M54/Sagittarius (purple), NGC~6273/Kraken (red), and NGC~6934/Helmi streams (blue). \ocen/GSE is in green. Black dotted lines represent different values for the \mnsc/\mstar ratio. 
\label{fig:mnsc}}
\end{figure}

There are other candidate NSCs in the MW. NGC~6273 (M19) and NGC~6934 have been proposed to be NSCs of Kraken/Heracles \citep{Kruijssen2020kraken, Horta2021} and Helmi streams \citep[][]{helmi1999} disrupted dwarfs, respectively, by \citet[][]{Pfeffer2021}. However, unlike \ocen/GSE and M54/Sagittarius, both NGC~6273 and NGC~6934 would be significant outliers in the \mnsc--\mstar relation if really related to these satellites (Figure \ref{fig:mnsc}). These associations could be incorrect, but these candidate NSCs might have also experienced substantial mass loss through tides. Future studies on the stripping history of NSCs in dwarf galaxies merging with massive hosts would be of interest to test if these systems would really be expected to be outliers in the \mnsc--\mstar relation.

With respect to the NSC candidates themselves, there is strong evidence from high-resolution stellar spectroscopy that NGC~6273 
is a genuine NSC \citep[][]{Johnson2017ngc6273_m19}. Hence, there might be a low-mass ($\mstar \lesssim 10^7\,\msun$) dwarf-galaxy host for NGC~6273 still undiscovered in the MW. On the other hand, confirmation of NGC~6934 as a true NSC is still needed; the best spectroscopic sample available is from \citet[][]{marino2021gcs} with only 13 stars. Other aforementioned candidate stripped NSCs in M31 with claims of IMBHs are G1 \citep[][$\mbh \sim 20{,}000\,\msun$]{Gebhardt2002g1} and B023-G078 \citep[][$\mbh \sim 100{,}000\,\msun$]{Pechetti2022imbh}. In Figure \ref{fig:relat}, I only provide lower limits on \mstar for these clusters since their original hosts are unknown. 

\section{SUMMARY} \label{sec:conclusions}


I have demonstrated that the proposed central IMBH inside \ocen/GSE follows the local (redshift $z \sim 0$) \mbh--\mstar relation. This suggests that this scaling relation might extend to GSE-mass, i.e., S/LMC-mass galaxies. Hence, since there are many Local Group satellites that could host central IMBHs, this result might encourage new observational programs to hunt for them. Sagittarius dSph could be a compelling target since it is relatively close and it also hosts the NSC M54. Further theoretical studies  could also be promising such as on the dynamical response of dwarf galaxies to central IMBHs and the AGN feedback mechanism for regulating the \mbh--\mstar relation in low-mass systems.


I have also shown that the IMBH in \ocen agrees with the \mbh--$\sigma_\star$ relation. Other stellar clusters with candidate IMBHs in M31, namely G1 and B023-G078, as well as UCDs also reside on top of this relation, perhaps corroborating the connection between stripped NSCs and these compact galaxies. Moreover, NSCs and UCDs might be comparable to bulges of local massive galaxies in the \mbh--$\sigma_\star$ relation. Other wandering accreted NSCs in the MW might also offer the opportunity to push further the \mbh--\mstar relation into the low-mass regime and inform about IMBH demographics. Confirming the host galaxy of NGC 6273 (M19) and if NGC 6934 is a genuine NSC are of particular urgency. 

I also performed an initial exploration of BH growth histories for \ocen/GSE. This analysis suggest that, if this IMBH formed from direct collapse, its initial mass must have been quite low compared to the expected distribution from heavy seeding prescriptions. The accretion rate of this IMBH must have also stayed very low throughout its $\sim$3\,Gyr lifetime until the GSE merger with the MW. This behavior could be the complete opposite of high-$z$ AGNs similar to GN-z11 identified to host overmassive BHs that require both high initial masses and accretion rates. Another possibility is that \ocen/GSE's IMBH formed from light seeding from a Population III supernova remnant. One implication could be that both heavy and light seeding mechanisms are at play to form central BHs, further reinforcing the importance of finding additional IMBHs. 



\begin{acknowledgments}
I sincerely thank the anonymous referee for excellent comments and suggestions that contributed to this manuscript. I mainly acknowledge funding from FAPESP (proc. 2021/10429-0) and KICP/UChicago through a KICP Postdoctoral Fellowship. I also thank Fabio Pacucci for providing code to calculate black hole growth trajectories and Vini Placco for the title suggestion. This paper was greatly inspired by discussions and presentations at the First Stars VII conference in New York City/USA. I am also indebted to all those involved with the multi-institutional \textit{Milky Way BR} Group for the weekly discussions. Finally, I am particularly grateful to Lais Borbolato, Silvia Rossi, Alex Ji, and Vini Placco (again) for encouraging me to actually write this paper which, otherwise, would end up being just a really long Twitter thread.

\clearpage




\end{acknowledgments}

\newpage
\setlength{\bibsep}{0pt}
\bibliographystyle{aasjournal}
\footnotesize

\bibliography{bibliography.bib}{}

\end{document}